\documentstyle[twoside,fleqn,espcrc2]{article}

\newcommand{\AmS}{{\protect\the\textfont2
  A\kern-.1667em\lower.5ex\hbox{M}\kern-.125emS}}

\newcommand{\bbone}{{\mathchoice {\rm 1\mskip-4mu l} {\rm 1\mskip-4mu l}
{\rm 1\mskip-4.5mu l} {\rm 1\mskip-5mu l}}}

\newcommand{\rvacp}{\vert + \rangle}
\newcommand{\rvacm}{\vert - \rangle}

\newcommand{\lvacp}{\langle + }
\newcommand{\lvacm}{\langle - }
 
\newcommand{\rVacp}{\vert v+ \rangle}
\newcommand{\rVacm}{\vert v- \rangle}
\newcommand{\lVacp}{\langle v+}
\newcommand{\lVacm}{\langle v-}

\title{
\vspace{-1cm}
\hfill
\parbox{6cm}{\normalsize RU-97-78\\ KUNS-1465 HE(TH)~97/14}\\
\vspace{0.5cm}
Gauge Freedom in Chiral Gauge Theory 
with Vacuum Overlap\thanks{Talk presented at Lattice '97, Edinburgh.}}

\author{Yoshio Kikukawa\address{Department of Physics and Astronomy, 
        Rutgers University, Piscataway, NJ 08855-0849, USA}%
        \thanks{On leave of absence from: Department of Physics, 
                Kyoto University, Kyoto 606-01, Japan}
       }
\begin{document}

\begin{abstract}
Dynamical nature of the gauge degrees of freedom and its effect
to fermion spectrum are studied at $\beta=\infty$ for two- and 
four-dimensional nonabelian chiral gauge theories in the vacuum 
overlap formalism. 
It is argued that the disordered gauge degrees of freedom does 
not contradict to the chiral spectrum of lattice fermion. 
\end{abstract}

\maketitle

\section{Introduction --- Pure gauge limit}
In the vacuum overlap formalism\cite{original-overlap} of a 
generic chiral gauge theory, 
gauge symmetry is explicitly broken by the complex phase of 
fermion determinant. In order to restore the gauge invariance,
gauge average ---the integration along gauge orbit--- is invoked.
Then, what is required for the dynamical nature of the 
gauge freedom at $\beta=\infty$ (pure gauge limit) is that 
the global gauge symmetry is not broken spontaneously and 
the bosonic fields of the gauge freedom have large mass compared 
to a typical mass scale of the theory so as to decouple from physical 
spectrum\cite{gauge-symmetry-restoration,original-overlap}.

However, through the analysis of the waveguide 
model\cite{original-waveguide,waveguide-analysis-golterman}, 
it has been claimed that this required disordered nature of the gauge 
freedom causes the vector-like spectrum of fermion\cite{overlap}. 
In this argument, the fermion correlation functions at 
the waveguide boundaries were examined. 
One may think of 
the counter parts of these correlation functions in the overlap 
formalism by putting creation and annihilation operators in 
the overlap of vacua with the same signature of mass. 
Let us refer this kind of correlation function as 
{\it boundary correlation function} and 
the correlation function in the original definition as 
{\it overlap correlation function}.
We should note that the boundary correlation functions 
are no more the observables in the sense defined in the overlap
formalism; they cannot be expressed by 
the overlap of two vacua with their phases fixed by the Wigner-Brillouin 
phase convention. But, they are still relevant because they can probe 
the auxiliary fermionic system for the definition of 
the complex phase of chiral determinant and therefore the 
anomaly (the Wess-Zumino-Witten term).
If massless chiral states could actually appear in the boundary 
correlation functions, we would have difficulty defining the 
complex phase. 

Our objective in this article is to argue that 
the disordered nature of the gauge degrees of freedom 
does not contradict to the chiral spectrum of lattice fermion
in the vacuum overlap formalism. 
For this goal, we consider the pure gauge limit 
of two- and four-dimensional $SU(N)$ nonablelian chiral gauge 
theories\cite{kikukawa-gauge-two-dimensions,kikukawa-gauge-four-dimensions}. 
The partition functions of these theories can be written in general as 
\begin{eqnarray}
    Z&=& \int [dU] \exp \left(-\beta S_G \right) \times \nonumber\\
     &&  \prod_{rep.} \left( 
        \frac{\lvacp \rVacp}{|\lvacp\rVacp|} 
               \lVacp \rVacm 
        \frac{ \lVacm \rvacm}{| \lVacm \rvacm|}  
              \right) .
\end{eqnarray}
In this formula\cite{original-overlap}, 
$\rVacp$ and $\rVacm$ are the vacua of the second-quantized 
Hamiltonians of the three- or five-dimensional Wilson fermion 
with positive and negative bare masses, respectively.
$\rvacp$ and $\rvacm$ are corresponding free vacua.
The Wigner-Brillouin phase convention is explicitly implemented 
by the overlaps of vacua with the same signature of mass. 
$\prod_{rep.}$ stands for the product over all Weyl fermion multiplets 
in an anomaly free representation. 
$S_G$ is the gauge action.

At $\beta =\infty$, 
the gauge link variable is given in the pure gauge form:
\begin{equation}
\label{eq:pure-gauge-link-variable}
  U_{n\mu}= g_n g_{n+\mu}^\dagger \quad  g^{}_n \in SU(N) .
\end{equation}
Then the model describes the gauge degrees of freedom 
coupled to fermion through the gauge non-invariant piece of the 
complex phase of chiral determinants. 
\begin{eqnarray}
    Z&=& \int [dg] 
      \prod_{rep.} \left( 
   \frac{\lvacp \vert \hat G \rvacp}{|\lvacp \vert \hat G \rvacp|}
   \lvacp \rvacm
   \frac{\lvacm \vert \hat G^\dagger \rvacm}
        {|\lvacm \vert \hat G^\dagger \rvacm|}  
              \right)  \nonumber\\
     &\equiv& \int d\mu[g] .
\end{eqnarray}
$\hat G$ is the operator of the gauge transformation given by:
\begin{equation}
\hat G 
= \exp \left( \hat a_n^{\dagger i} \{\log g\}_i{}^j \hat a_{n j} \right) .
\end{equation}

In this limit, one of the possible definitions of the 
{\it boundary correlation functions} is given as follows 
for the case of the negative mass and in the representation $r$:
\begin{eqnarray}
\langle \phi_n{}_i \phi^\dagger_m{}^j  \rangle_{-r}
&\equiv&\frac{1}{Z} \int d\mu[g] \times \\
&& \frac{\lvacm \vert \hat G^\dagger 
  \left\{ \hat a_n{}_i \hat a^\dagger_m{}^j  
         -\frac{1}{2} \delta_{nm} \delta_i^j \right\}
      \rvacm_r}
     {\lvacm \vert \hat G^\dagger \rvacm_r } .
\nonumber
\end{eqnarray}
Note that it transforms under the $SU(N)$ global gauge transformation as 
\begin{equation}
\langle \phi_n{}_i \phi^\dagger_m{}^j  \rangle_{-r}
\longrightarrow 
(g_0{}_i^s)   
\langle \phi_n{}_s \phi^\dagger_m{}^t  \rangle_{-r}
(g_0^\dagger{}_t^j) .
\end{equation}

\section{Asymptotically free pure gauge models}
As we can see from the pure gauge limit of the original theory, 
the gauge average is invoked without any weight for the gauge freedom 
except for the complex phase of the chiral determinant. 
This way of the gauge average is expected to keep the disordered 
nature of the gauge freedom.  
In order to examine the dynamical effect 
of the gauge average, however, it is desirable to have control over 
the fluctuation of the gauge freedom\cite{2d-wess-zumino-by-overlap} 
by deforming the weight of the gauge average as
\begin{equation}
d\mu[g] \rightarrow d\mu[g;K]= d\mu[g] \exp\left( - K S_g[g] \right) .
\end{equation}
For nonabelian gauge groups, as we will argue, this can be acheived 
without spoiling its disordered nature 
{\it by introducing an asymptotically free self-coupling 
of the gauge degrees of freedom}. 

In two-dimensions, we introduce the nearest-neighbor 
coupling\cite{kikukawa-gauge-two-dimensions}:
\begin{equation} 
K \sum_{n\mu} {\rm Tr}
\left( g_n^{} g_{n+\hat\mu}^\dagger + g_{n+\hat\mu}^{} g_n^\dagger
\right), \quad \left(K \equiv \frac{1}{\lambda^2}\right) .
\end{equation}
In four-dimensions, we consider the covariant gauge fixing term 
and the Faddeev-Popov determinant\cite{kikukawa-gauge-four-dimensions}, 
following Hata\cite{pure-gauge-model} 
(cf.  \cite{rome-approach,renormalizable-gauge-model,covariant-type-gauge-fixing-action}):
\begin{equation}
-\frac{1}{2\alpha} \sum_n \left( \bar \nabla_\mu \hat A_{n\mu} \right)^2 
- \sum_{nm} \bar c_n^a \hat M^{ab}_{nm} c_m^b \, (\alpha\equiv\lambda^2),  
\end{equation}
where 
\[
\label{eq:lattice-vector-potential}
\hat{A}_{n\mu}=
 \frac{1}{2 i} \left( U_{n\mu}-U^\dagger_{n\mu} \right) 
-\frac{1}{N} \bbone \, 
{\rm Tr} \, 
    \frac{1}{2 i} \left( U_{n\mu}-U^\dagger_{n\mu} \right) ,
\]
with the pure gauge link variable Eq.(\ref{eq:pure-gauge-link-variable}).
$\hat{M}^{ab}_{nm}$ is the lattice Faddeev-Popov operator.

We can show by the perturbation theory in $\lambda$ (background field
method) that both pure gauge models share the novel 
features of the two-dimensional nonlinear sigma model
{\it even in the presense of the imaginary action}. 
The model is renormalizable (at one-loop in four-dimensions) 
and $\lambda$ is asymptotically free. Severe infrared divergence 
occurs and it prevents local order parameters from 
emerging. 

Based on these dynamical features, we assume that the
global gauge symmetry does not break spontaneously 
for the entire region of $\lambda$ and the gauge freedom acquires 
mass $M_g$ dynamically through the dimensional transmutation. 
With this assumption, the decoupling of the gauge freedom could occur 
as $M_g \nearrow \frac{1}{a}$ in the limit $\lambda \nearrow \infty$. 

In two-dimensions, an important counter example occurs 
if the gauge anomaly does not cancel and the actual Wess-Zumino-Witten  
term appears in the complex action. This term causes 
an IR fixed point in the beta function of $\lambda $\cite{witt}.
It has been shown to be true also in the vacuum overlap 
formalism\cite{2d-wess-zumino-by-overlap}.
The fixed point theory is equivalent to the free massless fermion 
with the chiral $SU(N)$ symmetry\cite{witt} and this simply means 
the failure of the decoupling of the gauge degrees of freedom.
It is interesting to note the possibility that 
the nature of the pure gauge dynamics could distinguish the 
anomaly-free chiral gauge theory from anomalous ones. 

\section{Boundary correlation function}
The asymptotic freedom allows us to tame the gauge fluctuation 
by approaching the critical point of the gauge freedom. 
There we can invoke the 
spin wave approximation for the calculation of the 
invariant boundary correlation function
\begin{equation}
 \langle \phi_n{}_i \phi^\dagger_m{}^i  \rangle_{-r} ,   
\end{equation}
which is free from the IR divergence. At one-loop order, we obtain
\begin{eqnarray}
&& \frac{1}{2} \delta_{nm} \delta_i^i - S^v_-( n-m ) \delta_i^i 
\nonumber\\
&-&
\lambda^2  \sum_{r} 
S^v_-( n-r ) 
\left[ 
\langle \pi_r \pi_m \rangle^\prime S^v_-( r-m ) 
\right] \delta_i^i 
\nonumber\\
&+&
\lambda^2  \sum_{r,l} 
S^v_-( n-r ) \times 
\\
&&
\left[ 
\langle \pi_r \pi_l \rangle^\prime S^v_-( r-l ) 
\right] S^v_-( l-m ) \delta_i^i  +{\cal O}(\lambda^4) ,
\nonumber
\label{eq:boundary-correlation-function-colored}
\end{eqnarray}
where 
\begin{eqnarray}
S^v_-(n-m)
&=&
\int \frac{d^D p}{(2\pi)^D} e^{i p (n-m)} \, \frac{1}{2 \lambda_-}
\times
\\
&& 
\left( \begin{array}{cc} 
-m_0+B(p) & \sigma_\mu \sin p_\mu \\
\sigma_\mu^\dagger \sin p_\mu & m_0-B(p)
       \end{array}\right) , \nonumber
\end{eqnarray}
\begin{eqnarray}
B(p)&=& \sum_\mu \left( 1-\cos p_\mu \right) , \\
\lambda_- &=& {\sqrt{\sum_\mu \sin^2 p_\mu+(B(p)-m_0)^2}}, 
\end{eqnarray}
and 
\begin{eqnarray}
\langle \pi_n \pi_m \rangle^\prime 
&=& \int \frac{d^D p}{(2\pi)^D} 
\frac{ e^{i p(n-m) } -1 }
{\left( \sum_\mu 4 \sin^2 \frac{p_\mu}{2} \right)^{D/2}}.
\end{eqnarray}
It turns out that the spectrum of $S^v_-$ has 
the mass gap $M_B$ of the order of the cutoff, 
\begin{equation}
\cosh M_B = 1 + \frac{ m_0^2}{2 (1 - m_0)}, \quad (m_0 = 0.5).
\end{equation}
We also see that the quantum correction due to the gauge
fluctuation at one-loop does not affect the leading short-distance nature. 
There is no symmetry against the spectrum mass gap.  
Therefore it seems quite reasonable to assume that it holds true 
as $\lambda$ becomes large. 
Since the overlap correlation function does not depend on the 
gauge freedom and does show the chiral spectrum\cite{original-overlap},
the above fact means that the entire fermion spectrum is chiral. 

The author would like to thank H.~Neuberger and R.~Narayanan for
enlightening discussions. He also would like to thank H.~Hata, S.~Aoki, 
H.~So and A.~Yamada for discussions.

\end{document}